\documentclass{aa}  

\usepackage{graphicx}
\usepackage{txfonts}
\usepackage{amsmath}	
\usepackage{amssymb}	
\usepackage{xspace}
\usepackage{enumitem}
\usepackage{natbib}
\bibpunct{(}{)}{;}{a}{}{,} 

\usepackage[hidelinks]{hyperref}

\newcommand{\package}[1]{\textsl{#1}}
\newcommand{\acronym}[1]{{\small{#1}}}
\newcommand{\gaia}{\textsl{Gaia}\xspace}

\newcommand{\DRthree}{\acronym{EDR3}\xspace}

\newcommand{\dhelio}{\ensuremath{\textrm{d}_\odot}\xspace}
\newcommand{\vect}[1]{\ensuremath{\mathbf{#1}}}

\newcommand{\kpc}{\ensuremath{\textrm{kpc}}\xspace}
\newcommand{\kms}{\ensuremath{\textrm{km}~\textrm{s}^{-1}}\xspace}

\newcommand{\Vgsr}{\ensuremath{\textrm{V}_\textrm{GSR}}\xspace}
\newcommand{\Vlsr}{\vect{V}\ensuremath{_\textrm{LSR}}\xspace}
\newcommand{\Vvec}{\vect{V}\xspace}
\newcommand{\poe}{\ensuremath{\varpi/\sigma_{\varpi}}\xspace}
\newcommand{\Lz}{\ensuremath{L_{z}}\xspace}
\newcommand{\Lperp}{\ensuremath{L_{\perp}}\xspace}
\newcommand{\Etot}{\ensuremath{E_\textrm{tot}}\xspace}

\newcommand{\rev}[1]{#1}

\begin{document}

   \title{Linking nearby stellar streams to more distant halo overdensities}

   \author{E. Balbinot
          \and
          A. Helmi
          }

   \institute{Kapteyn Astronomical Institute, University of Groningen, Postbus 800, 
              NL-9700AV Groningen, The Netherlands}

   \date{Received \today; accepted \today}
   
  \abstract
   {It has been recently shown that the halo near the Sun contains several
   kinematic substructures associated to past accretion events. For the more
   distant halo, there is evidence of large-scale density variations -- in the
   form of stellar clouds or overdensities.}
   {We study the link between the local halo kinematic groups and three of
   these stellar clouds: the Hercules-Aquila cloud, the Virgo Overdensity, and
   the Eridanus-Phoenix overdensity.}
   {We perform orbital integrations in a standard Milky Way potential of a
   local halo sample extracted from \gaia \DRthree, with the goal of
   predicting the location of the merger debris elsewhere in the Galaxy. We
   specifically focus on the regions occupied by the three stellar clouds
   and compare their kinematic and distance distributions with those predicted
   from the orbits of the nearby debris.}
   {We find that the local halo substructures have families of orbits that tend
   to pile up in the regions where the stellar clouds have been found. The
   distances and velocities of the cloud's member stars are in good
   agreement with those predicted from the orbit integrations, particularly
   for Gaia-Enceladus stars. This is the dominant contributor of all three
   overdensities, with a minor part stemming from the Helmi streams and to
   an even smaller extent from Sequoia. The orbital integrations predict no
   asymmetries in the sky distribution of halo stars, and they pinpoint where
   additional debris associated with the local halo substructures may be
   located.}
   {} 

   \maketitle

\section{Introduction}

The importance of accretion in the build up of the Galactic stellar halo was
made evident by wide-field photometric surveys such as the Sloan Digital Sky
Survey \citep[SDSS;][]{Yanny00, Ivezic12}. This revealed a plethora of
overdensities in the sky distribution of distant stars such as the Sagittarius
streams. Other notorius examples are the Virgo Overdensity and Hercules-Aquila
Cloud \citep[VOD and HAC;][]{Vivas01, Newberg02, Belokurov07, Juric08,
Bonaca12, Conroy18}; and the more recently discovered Eridanus-Phoenix
Overdensity \citep[EriPhe;][]{Li16}. These three structures are sparse, span
several hundreds of deg$^2$ and have similar galactocentric distances of $\sim
15-20$~kpc.

The \rev{photometrically identified} overdensities unambiguously demonstrate
that the outer halo of the Milky Way was built via mergers and accretion. For
the inner halo, the evidence was less compelling until the advent of large
samples of stars with 6D phase-space information, particularly from \gaia DR2
\citep{Gaia2018}. This dataset revolutionised our view of the nearby halo, and
revealed that this is dominated by debris from \gaia-Enceladus (a.k.a.~the
Sausage; hereafter G-E) \rev{\citep{Helmi18, Belokurov18, Haywood18,
Mackereth19, Myeong19}}, and from the disk that was present at the time and was
dynamically perturbed \rev{\citep[][see also \citealt{Haywood18, Mackereth19,
Myeong19, Grand20, Belokurov20}]{Helmi18, Gallart19, DiMatteo2019}}.Besides
these major contributors, smaller kinematic substructures have been identified,
such as the Helmi streams \citep[HStr hereafter;][]{Helmi99, Helmer19b},
Thamnos \citep{Helmer19a}, Sequoia \citep{Myeong19} and several others
\citep[see e.g.][]{Naidu20}. 

It is natural to ask what the link is between the overdensities identified at
large distances, and the substructures nearby. For example, \citet{Deason18}
has argued that the break in the counts of halo stars at $\sim 20-25$~kpc
\citep{Deason13} corresponds to the piling up of stars from {G-E} at the
apocentre of their orbits. \citet{Simion18, Simion19} have used the distances,
line-of-sight velocities and proper motion information of stars associated to
the VOD and HAC to argue that these are likely the distant counterparts of
\gaia-Enceladus, because of their radially elongated orbits.  \cite{Donlon19,
Donlon20, Naidu21} suggest a similar association based on N-body simulations,
although the merger time inferred by \citet{Donlon20} does not match that
estimated for \mbox{G-E} using nearby stars \citep{Helmi18,Chaplin20}. Also
EriPhe \citep{Li16, Donlon19, Donlon20}, has been linked to those overdensities
because of its similar heliocentric distance. Here we take a complementary
approach, and integrate the orbits of halo stars in the solar vicinity
associated to various kinematic groups to predict where other stars with a
different orbital phase, but following similar orbits (as expected if they
originate from the same progenitor), would be located. We also compare their
proper motions and line-of-sight velocities where available to those predicted
by the orbit integrations. As we show in this paper, although many of the stars
in the overdensities can be linked to G-E, this is likely not the only
contributor.

This paper is organised as follows. In Sec.~\ref{sec:data} we define our local
halo sample and a set of distant tracers compiled from literature. We also
describe the framework which we use to integrate the orbits of nearby halo
stars. In Sec.~\ref{sec:results} we explore the overlap, in observable space,
of the nearby halo substructures' orbits with known stellar over-densities.
Finally, in Sec. \ref{sec:discussion} we discuss and summarise our findings.

\section{Data and Methods}
\label{sec:data}

\subsection{Nearby halo sample, streams and orbit integration}
\label{sec:sample}

We build a halo sample from the \gaia EDR3 6D subset \citep{Gaiaedr3}, with a
quality cut on parallax of $\poe > 5$, and with distances computed by inverting
the parallaxes after applying a zero-point correction of $0.017\,\mu$as as
recommended by \citet{GaiaLindegren}. To obtain Galactocentric velocities, we
assume $v_{\rm LSR}$ = $232.8\, \kms$ \citep{McMillan17}, and a peculiar motion
for the Sun of $(U_\odot,V_\odot,W_\odot) = (11.1, 12.24, 7.25)$~km/s 
\citep{Schoenrich10}. We also assume the Sun is located at $R$ = $8.2$~\kpc
from the Galactic centre, and $z$ = $20.8$~pc above the plane
\citep{Bennett19}. We identify as nearby halo-like stars those with distances
\dhelio~$<$~2.5~\kpc, and that satisfy |\Vvec - \Vlsr|~$>$~210~km/s. This
selects 20010~stars.

For this halo sample, we compute their total energy \Etot, the angular momentum
in the $z$-direction $L_z$, (which we set to be positive in the direction of
Galactic rotation), and the circularity $\eta = L_z/L_z(\Etot)$, where
$L_z(\Etot)$ is the angular momentum of a circular orbit in the Galactic plane
with energy $\Etot$ \citep[see][]{Wetzel11}. We assume the \texttt{MWPotential}
from the dynamics package \texttt{gala}\footnote{\url{http://gala.adrian.pw/}} \citep{gala}. This Galactic potential
has a spherical nucleus and bulge, a Miyamoto-Nagai disk \citep[with parameters
from][]{Bovy15}, and a spherical NFW dark matter halo, where the values of the
parameters have been obtained by fitting to a compilation of mass measurements
of the Milky Way, from 10~pc to $\sim 150$~kpc. 

We use \Etot, \Lz and $\eta$ to isolate four substructures following the work
of \citet{Helmer19a}, namely \gaia-Enceladus, Sequoia, the Helmi streams and
Thamnos. These authors split Thamnos into two groups, although for our analysis
we will treat them as one. In the top panel of Figure~\ref{fig:ELz} we show
their distribution in \Lz--\Etot space, with the various substructures
colour-coded. The selection criteria adopted for each of them are:
\begin{description}[font=$\bullet~$\scshape\bfseries]
    \item[G-E:] $-1.30 < \Etot/u_E < -0.70$ and $ -0.20 < \eta < 0.13$;
    \item[Helmi streams:] $1.6 < \Lperp/u_L < 3.2$ (where $\Lperp \equiv ({L_x^2} + {L_y^2})^\frac{1}{2}$) and $1.0 < \Lz/u_L < 1.5$  and $\Etot/u_E < -0.9$;
    \item[Sequoia:]   $ -1.10 < \Etot/u_E < -0.90 $ and $ -0.65 < \eta < -0.35 $
        and $\Lz/u_L > -2.1$;
    \item[Thamnos 1:] $ -1.00 < \Etot/u_E < -0.75 $ and $\eta < -0.75 $, 
    \item[Thamnos 2:] 
    $ -1.52 < \Etot/u_E < -1.40 $ and $-0.75 < \eta < -0.4,$ 
\end{description}
where $u_E = 10^5$~km$^2$/s$^2$, and $u_L = 10^3$~kpc km/s. 

We integrate the orbits of the stars in our sample backwards in time for 8 Gyr
with a time sampling of 10 Myr in the same Galactic potential using a
\citet{DOPRI} integrator. We refer to the time-sampled positions and velocities
as \emph{orbital points}. We expect in this way to pinpoint the location of
other stars sharing the same progenitor as the substructures mentioned above,
but having a different orbital phase. 

\begin{figure}
\centering
   \includegraphics[width=0.45\textwidth]{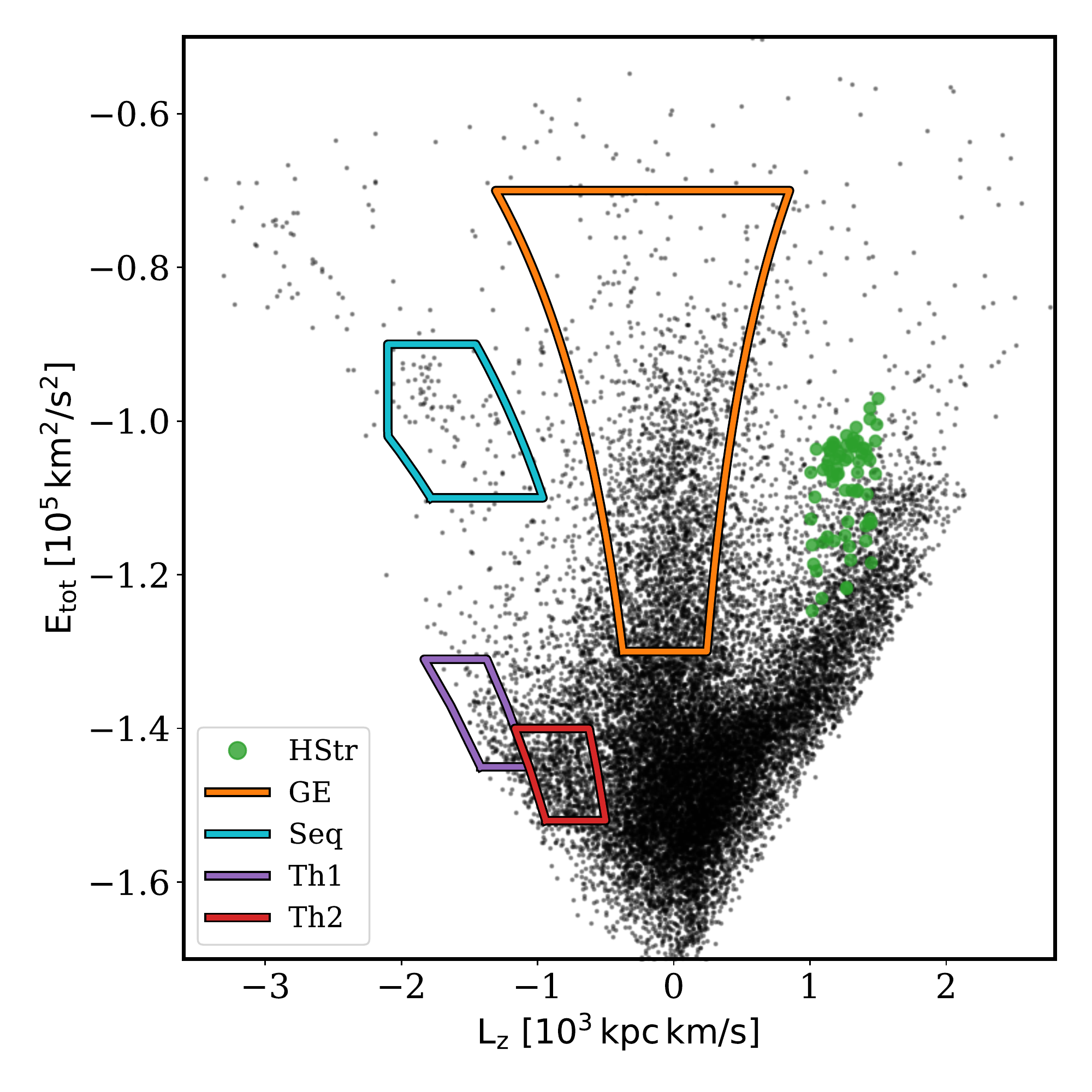}
   \includegraphics[width=0.45\textwidth]{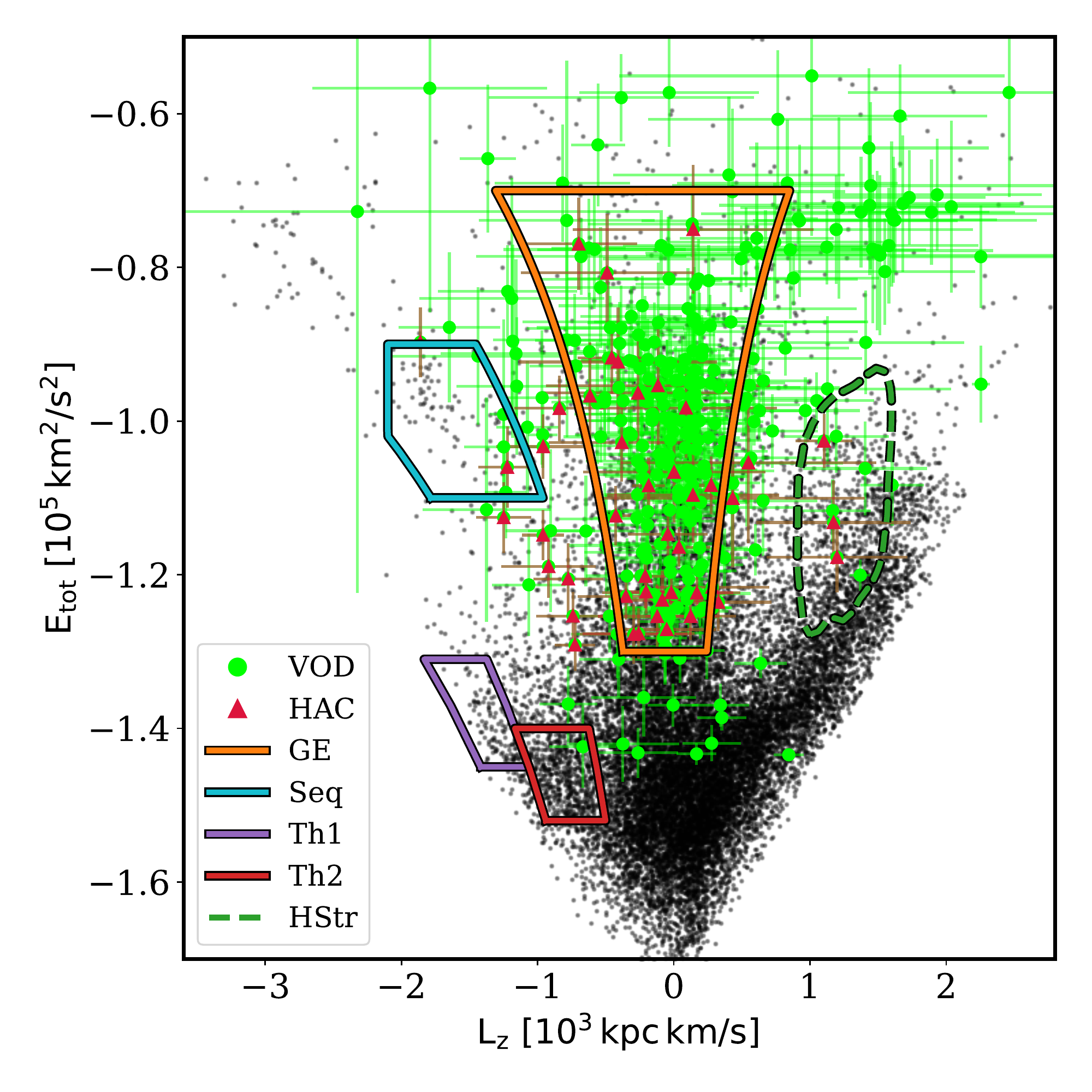}
   \label{fig:ELz}
   \caption{\emph{Top panel:} \Lz--\Etot for the stars in our nearby halo
       sample from \gaia EDR3, where we have marked the different structures
       studied in this paper. \emph{Bottom panel:} same as the top panel, but
       we now show the energy and angular momentum for the distant tracers
       associated with VOD and HAC.  Estimates for the uncertainty in \Lz and
       \Etot are given as errorbars for the distant tracers only (as they are
       large). Note that the Helmi streams (HStr) are mostly selected in
       another subspace, namely $L_z$--$L_\perp$, so we display (the region
       occupied by) its members differently. }
\label{fig:ELz}
\end{figure}

\begin{figure*}
\centering
   \includegraphics[width=0.9\textwidth]{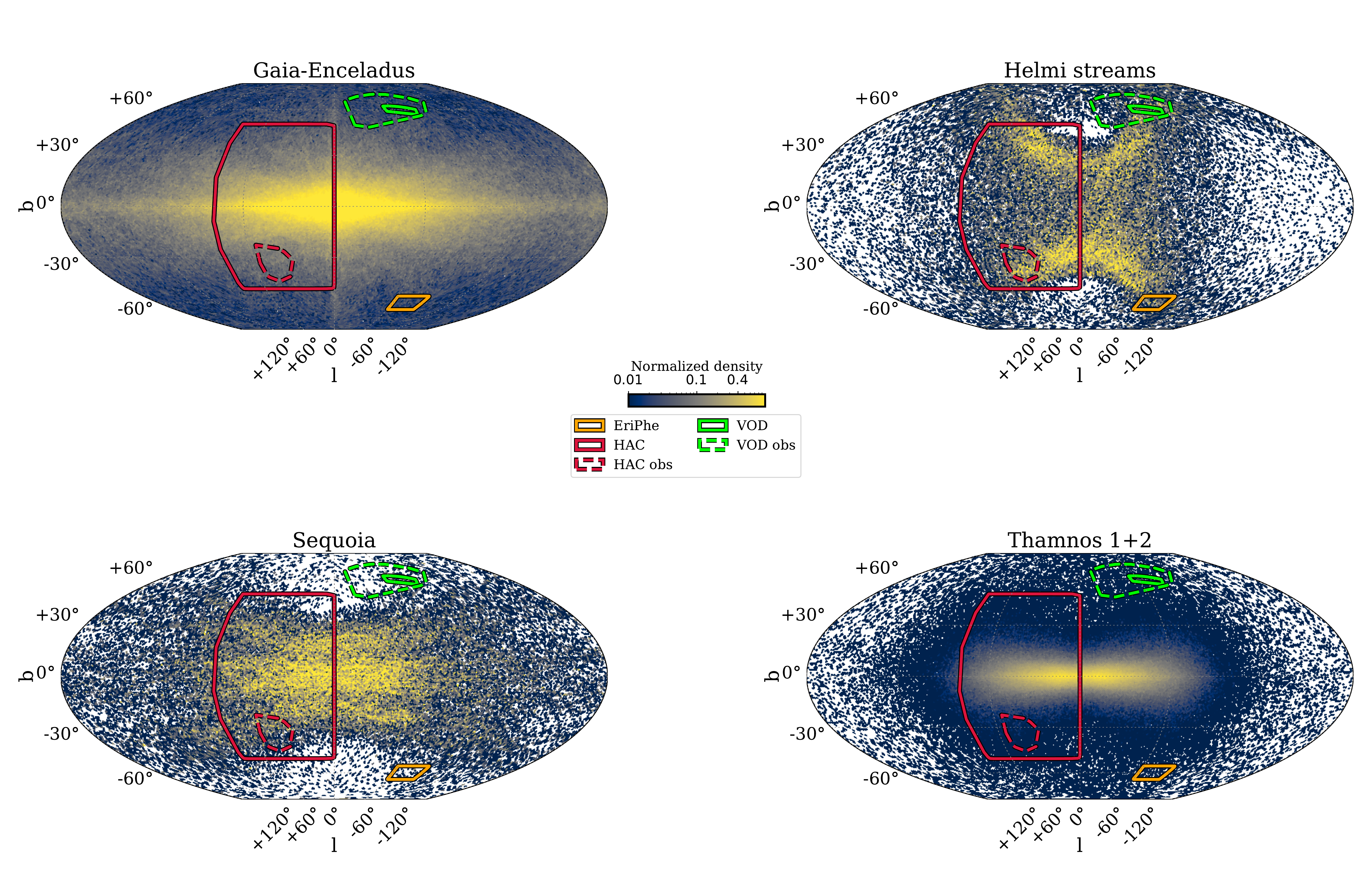}
     \caption{McBryde--Thomas projection of the density of orbital points for
         the different nearby halo substructures (indicated on top of each
         panel). The colormap represents high values as yellow and low as blue,
         with the density normalised per panel. The reported
         \texttt{galstreams} locations of known overdensities are shown as
         polygons, where the dashed polygons indicate the position of spectroscopic
        confirmed members (when available).}
     \label{fig:allsky}
\end{figure*}

\subsection{Distant tracers and overdensities}
\label{sec:clouds}

We compile a set of distant tracers that have reliable distance estimates that
go beyond the \gaia EDR3 6D sample. For the VOD we take the combined sample of
\citet{Vivas16} and \citet{Sesar17} that contains distances and radial velocity
measurements for RR Lyrae stars. We note that this sample also contains stars
associated to the Sagittarius stream, which can be filtered out by selecting
only stars with \dhelio $< 40$ \kpc. 

For the HAC we use the sample from \citet{Simion18} which comprises RR Lyrae
derived distances plus radial velocities for 45 stars. This sample concentrates
on a small portion of the reported extent of the HAC
\citep[see][]{Grillmair16}. 

Due to its recent discovery, the EriPhe overdensity \citep{Li16} has not had
any extensive spectroscopic follow-up, \rev{implying that there are no radial
velocity measurements}.  Here we define a list of possible members from the
Dark Energy Survey (DES) Year 6 RR Lyrae catalogue \citep{DESRRLyY6}.  We find
41 stars in the reported region and distance range for EriPhe \citep{Li16}.
The stars in the VOD and HAC samples were positionally cross-matched to \gaia
EDR3 astrometry, while the EriPhe was cross-matched using the DES-provided
\gaia DR2 \texttt{source\_id}, which was transformed to \gaia EDR3 \texttt{id}
via the \texttt{dr2\_neighbourhood} table, where the closest angular distance
neighbour was selected as the best-neighbour.

In the bottom panel of Figure \ref{fig:ELz} we show the \Lz-\Etot distribution
of distant tracers associated with the stellar clouds discussed above. We
assume a 10\% uncertainty  in distances for the RR Lyrae. The radial velocity
uncertainties are provided by each catalogue and range from 5-20~km/s.  We
consider the correlation of uncertainties in $\mu_{\alpha*}$ and
$\mu_{\delta}$, while assuming the uncertainty in the on-sky position to be
negligible. Although the total uncertainties are sizeable, we note from
Fig.~\ref{fig:ELz} that many of the stars, particularly those from the VOD
appear to be associated to G-E \citep[as previously reported by][on the basis
of their angular momenta]{Simion19}. 

We compute the reported extent of the overdensities on the sky in Galactic
coordinates using the \texttt{galstreams} package \citep{galstreams}, as shown
in  Figure \ref{fig:allsky}  with solid lines. For each of samples described
above, we also defined a region on the sky (in dashed) based on the convex hull
of their observed members. We note that for the HAC the locus of
spectroscopically confirmed members (in dashed) is smaller than the full extent
of the cloud. The opposite is true for VOD due to the inclusion of the more
recent samples from \citet{Vivas16, Sesar17}. 

\section{Results}
\label{sec:results}

We now compare the spatial and kinematic characteristics of the orbits of the
nearby halo substructures identified in \Lz-\Etot (Sec. \ref{sec:sample}) in
the regions of the sky occupied by the more distant overdensities (Sec.
\ref{sec:clouds}). The hypothesis behind this comparison is that stars from a
given progenitor have similar orbits, and that by selecting the stars near the
Sun we sample objects with a specific orbital phase, i.e. those that happen to
pass currently through the Solar volume. We thus expect other stars from that
same progenitor to be located presently elsewhere in the Galaxy, along the
similar orbital tracks but with different orbital phases. 

\begin{figure*}[ht!]
\centering
   \includegraphics[width=\textwidth]{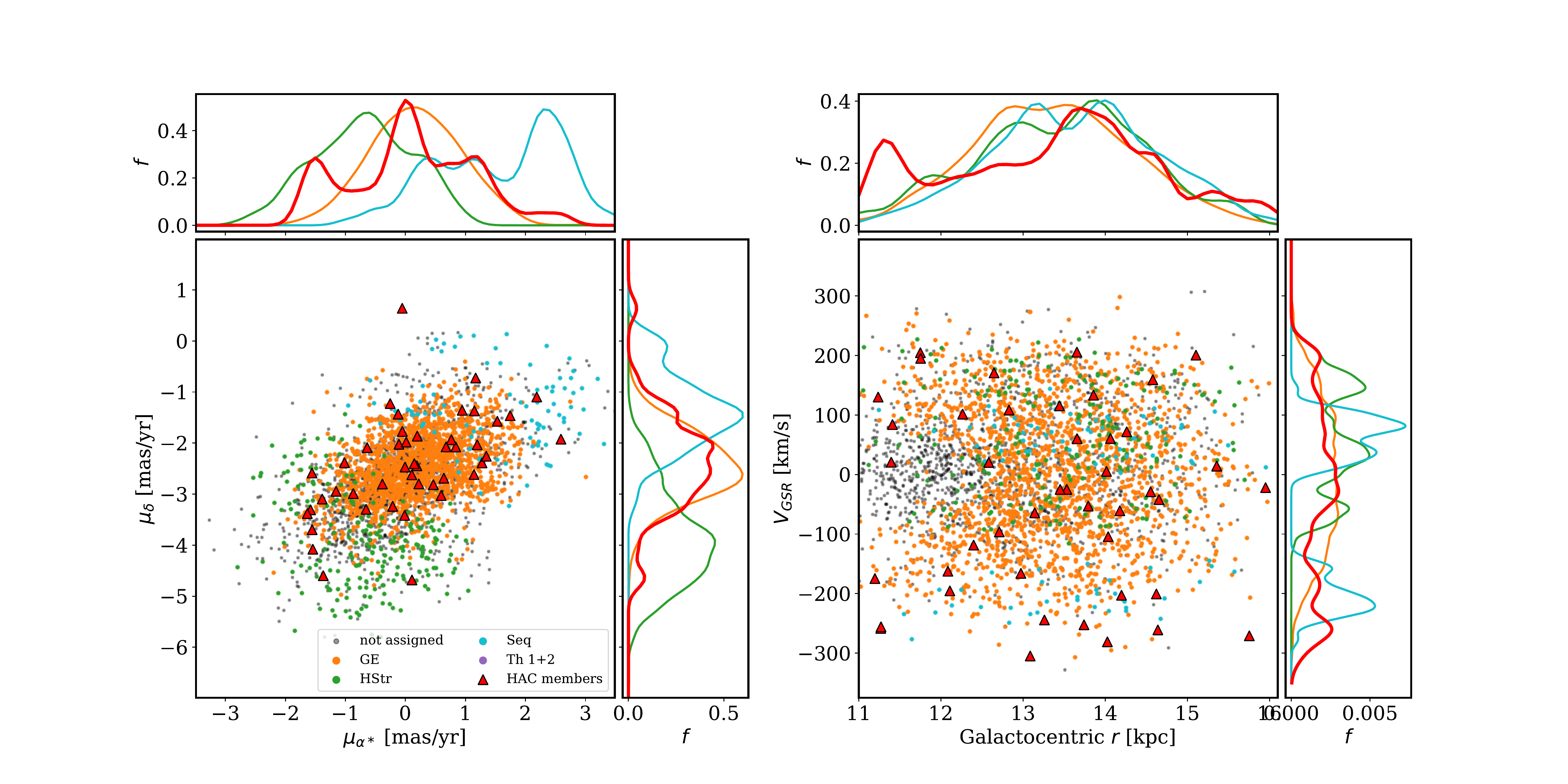}
     \caption{\emph{Left panel:} PM distribution of orbital points derived from
         the integration of the trajectories of stars associated with nearby halo
         substructures, with each colour corresponding to a different object
         (see legend).  The larger red triangles are observed members of the
         HAC. The smaller vertical and horizontal diagrams show the
         (normalized) KDE (bandwidth=0.2~mas/yr) for each substructure and for
         the observed HAC members. \emph{Right panel:} As in the left panel
         but showing the Galactocentric distance vs.  \Vgsr. The KDEs were
     computed using a 0.5 \kpc and 10~km/s bandwidth.  }
     \label{fig:HAC}
\end{figure*}

In Figure \ref{fig:allsky} we show an all-sky
\textsc{HEALpix}\footnote{\url{http://healpix.sourceforge.net}} map of the
density orbital points for the 4 nearby substructures mentioned in Sec.
\ref{sec:sample}. The density is normalized for each panel separately and
displayed in $\log_{10}$ scale, and serves to indicate where we might expect to
find contributions from the different substructures. The sky distributions for
G-E and Sequoia cover a large range in latitudes ($|b| \lesssim 60^\circ$) but
reveal as well the presence of stars on low inclination orbits, which are
presumably due to "contaminants" (that follow a very different orbital family).

Fig.~\ref{fig:allsky} shows that  the position on the sky of the VOD, HAC, and
EriPhe overdensities, 
coincide with regions with a high density of orbital points, 
\rev{from at least one of the halo substructures}, which explains why
the surveys have detected them in the first place. The occurrence of such
stellar clouds is also related to the availability of high-quality homogeneous
photometric data in low dust extinction regions. Given the lack of coverage in
some regions, namely between -60$^\circ < b < -30^\circ$ and 280$^\circ < l <
320^\circ$ we would predict these to hold undiscovered debris. This region is
also occupied by the Magellanic Clouds, which would make the detection of a
sparse stellar cloud very difficult without extensive spectroscopic data. The
``missing" stellar clouds would link the VOD -- which is known to extend down to
$b \sim 30^\circ$ in the northern galactic cap \citep{Bonaca12, Conroy18} -- to
EriPhe, in a similar way as HAC has been detected on both sides of the disk.

We now explore the distribution of orbital points in the spaces of proper
motion (PM), Galactocentric distance ($r$), and line-of-sight velocities
corrected for the Galactic Standard of Rest frame (\Vgsr). Figure \ref{fig:HAC}
shows the distribution of orbital points and observed members of the HAC in two
of these spaces. In PM we observe that the HAC shows a distribution that peaks
at the expected value for G-E, and also overlaps with the HStr. The
Galactocentric distance range for Thamnos is not consistent with that of the
HAC members, although note that the sample of \citet{Simion18} was selected to
have $15 <$ \dhelio/kpc $<20$. Furthermore, although the orbital points from
Sequoia match the tail of the \Vgsr distribution, these points do not match the
PM measured for the HAC stars. The most negative radial velocities of stars in
HAC appear to be related to G-E stars that have their orbital apocentre
well-beyond the distance range selected by \citet{Simion18}. 

\begin{figure*}[ht!]
\centering
   \includegraphics[width=\textwidth]{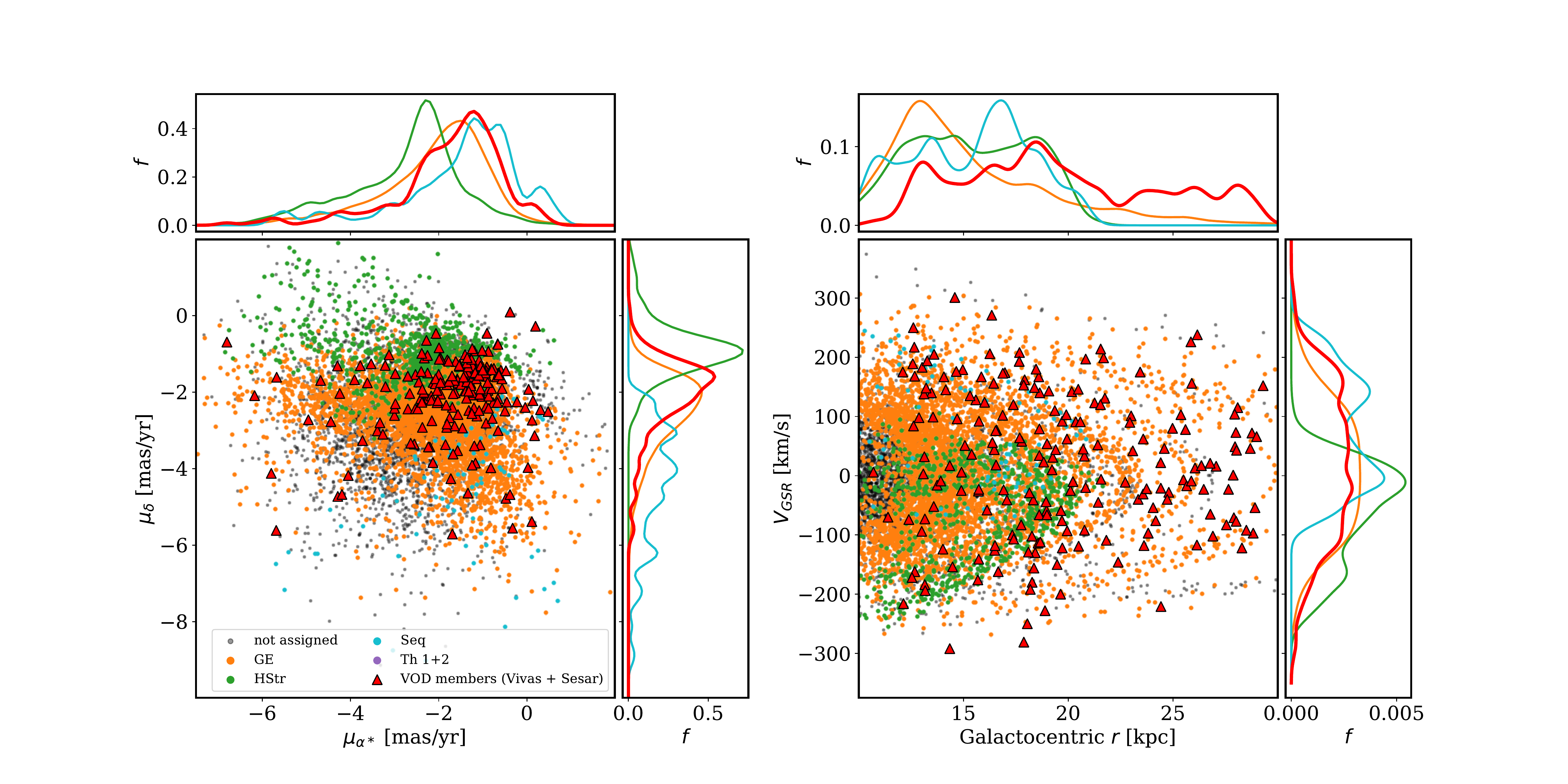}
     \caption{Same as Figure \ref{fig:HAC}, but now showing a comparison in the
     region of the sky occupied by the VOD. In the \emph{left} panel we show
     VOD members from \citet{Vivas16} and \citet{Sesar17}, while in the
     \emph{right} panel only the set from \citet{Vivas16} with measured \Vgsr is
     shown.}
     \label{fig:VOD}
\end{figure*}

Figure \ref{fig:VOD} shows the same analysis as above but now for the VOD. We
observe that its distribution in PM seems to be consistent with a combination
of contributions, mostly from the HStr and G-E, and to some extent also Sequoia
as there is one group of stars with $\mu_\delta <-3$~mas/yr and positive \Vgsr,
that matches also in distance.  The stars that overlap with the HStr orbital
points in PM, have distances and \Vgsr that confirm an association to this
object. Nonetheless, the majority of the VOD stars follow most closely the
distribution of the orbital points for G-E. 

\begin{figure*}[ht!]
\centering
    \includegraphics[width=\textwidth]{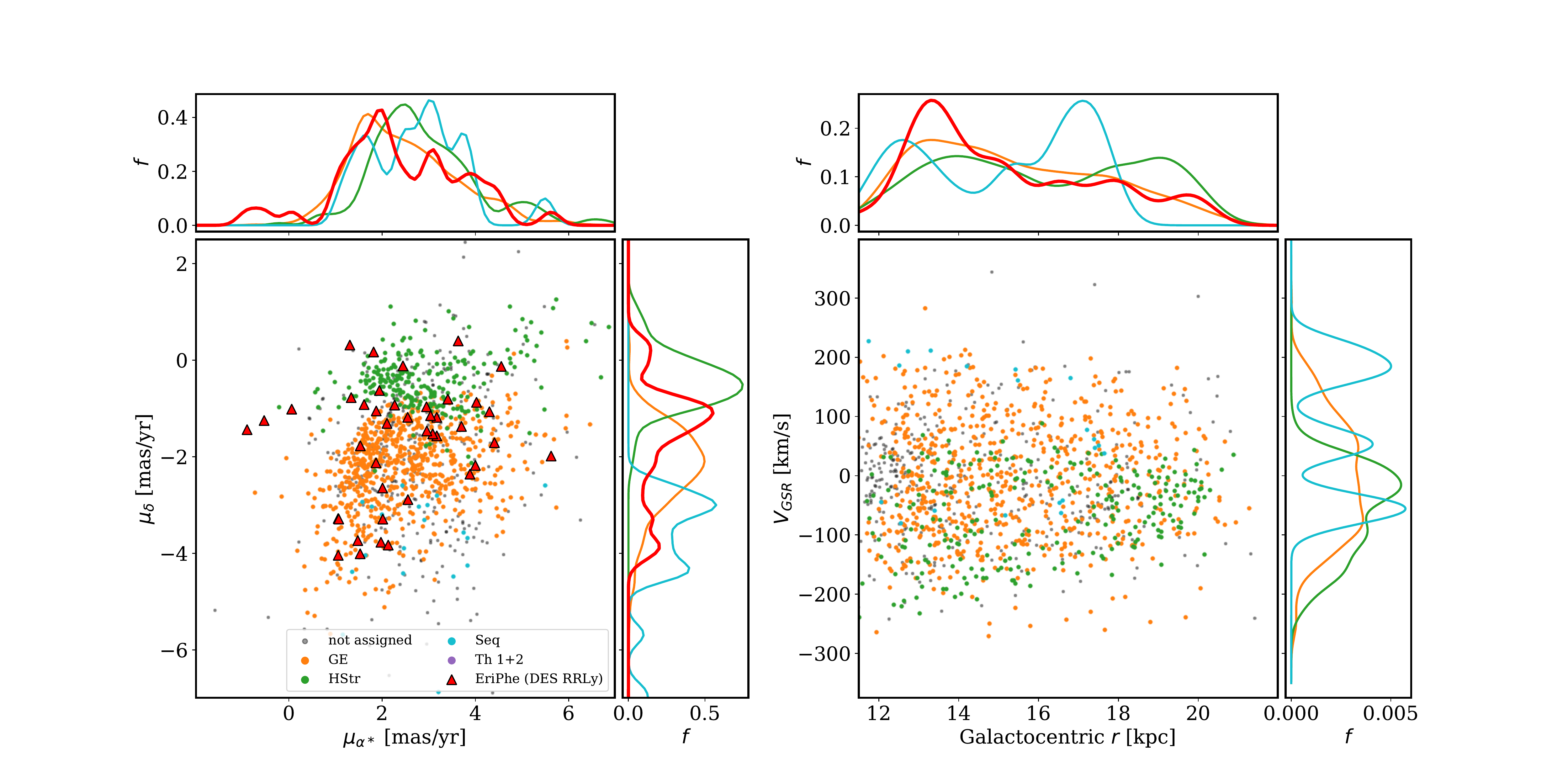}
     \caption{Same as Figure \ref{fig:HAC}, but now for the EriPhe overdensity.
     Note that no line-of-sight velocities are available for its members, so
     the right panel depicts a prediction of what to expect if EriPhe would have a 
     progenitor in common with the nearby halo substructures described in Sec.~\ref{sec:data}.}
     \label{fig:EP}
\end{figure*}

\begin{figure}[ht!]
\centering
    \includegraphics[width=0.40\textwidth]{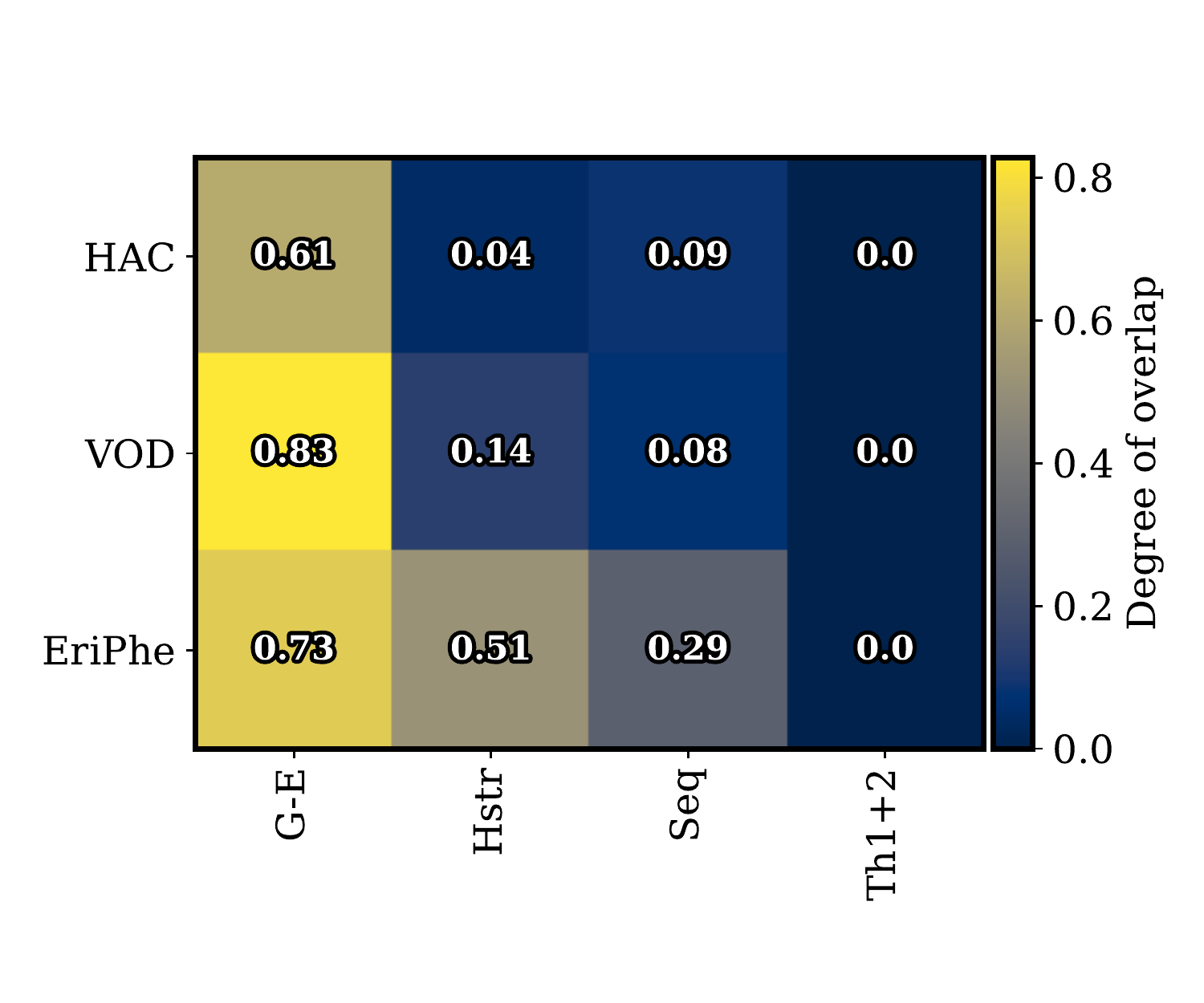}
     \caption{Matrix showing the degree of overlap in kinematic space
         namely $(\mu_\alpha, \mu_\delta, \Vgsr)$ for HAC and the VOD, and in
         $(\mu_\alpha, \mu_\delta)$ for EriPhe in the region on the sky where
         these data are available compared to the location of 95\% of the
         points resulting from the orbital integration.}
     \label{fig:overlap}
\end{figure}
\section{Discussion and Conclusions}
\label{sec:discussion}

Finally, in Figure \ref{fig:EP} we present the comparison for EriPhe.  This
stellar cloud has limited data, and its members were selected based solely on
their distances. We observe however, that the distribution in PM seems to match
quite well the predicted contributions of G-E and the HStr. While there is some
overlap in PM with Sequoia's orbital points, this is for less than a handful
EriPhe stars (with $-4 < \mu_\delta < -3$~mas/yr, at distances 16-17.5 kpc) and
on average EriPhe members tend to have smaller $r$ compared to what is expected
from Sequoia's orbital points. Just like for the VOD, EriPhe does not seem to
be linked to Thamnos. 

\rev{To better assess the association of clouds to halo substructures we devise
a test to map the degree of overlap in velocity space of these objects. At
the region of the sky occupied by a cloud, we compute a kernel density
estimation for the orbital points in $(\mu_\alpha, \mu_\delta, \Vgsr)$
space, with the exception of EriPhe, where only proper motions are
available.  We determine volume in velocity-space where 95\% of the orbital
points lie. Finally, we count how many observed data points fall within
this volume. In Figure \ref{fig:overlap} we show the degree of overlap for
each cloud-substructure pair.  We find the degree of overlap to be
consistent with our qualitative analysis, indicating a stronger link
between GE and all the clouds, while also hinting at a connection between VOD
and EriPhe with the HStr.}

On the basis of the very good match between the PM, distances and radial
velocities of HAC members to orbital integrations of nearby members of G-E
debris \rev{we argue that they likely stem from this progenitor. Some minor
overlap is present with Hstr and Sequoia, however, only detailed chemical
abundances could reveal the relative contributions of these accreted objects in
the region of the Galaxy occupied by the HAC, and put constraints on their
(internal) properties.}

The link between the majority of the stars in the VOD and nearby G-E debris, is
similarly also very clear. In particular, many of the VOD members appear to be
associated with streams of stars wrapping around in their orbits, and
encompassing debris coming from and going towards their orbital apocentres.
Some, but not all, of the groups identified by \citet{Vivas16}
on the basis of distances and radial velocities seem to correspond to different
wraps of debris (some others do not, because their measured proper motions are
not consistent with clustering in phase-space). A small fraction of the stars
in the VOD appears to be related to the Helmi streams. This is to some extent
also apparent from the sky map of the orbit integrations of VOD member stars
shown in \citet[][their Fig.~6, central panel]{Simion19}.

Finally, EriPhe could be a mixture of debris from the Helmi streams and G-E, as
both again are expected to contribute in the region of the sky where this
structure is located. The orbital points obtained from the integration of the
trajectories of nearby stars from these objects jointly reproduce the proper
motions measured for stars in EriPhe. To confirm this and distinguish more
clearly to which progenitor they might belong, line-of-sight velocity
information as well as chemical abundances would be very helpful. The
contribution of Sequoia to both for EriPhe and the VOD appears to be marginal,
while an association to Thamnos can be ruled out on the basis of very different
distance distributions.

Our study confirms earlier suggestions that the 3 overdensities are related,
and shows how in a standard Galactic potential (in our case, the
\texttt{MWPotential} from provided by the {\tt gala} package) it is possible to
match qualitatively the orbital distributions of stars from nearby
substructures to those of distant overdensities. This lends support to their
true association; the next step is to combine the distant and nearby tracers to
constrain the properties of the Galactic potential in detail, using methods
such as made-to-measure \citep{Hunt2014} or Schwarzschild's orbital
superposition \citep[as in][]{Magorrian2019}, the latter in essence being
similar to the approach presented in this paper. 

Our orbital integrations are allowing us to pinpoint the regions of the Galaxy
most likely to be occupied by different halo substructures and can thus be used
to direct spectroscopic follow-up efforts to most efficiently probe them. They
also show that after 8 Gyr of integration and starting from the phase-space
distribution of nearby stars, we expect a predominantly symmetric distribution
of stars on the sky, both with respect to the Galactic plane as well as to the
Galactic centre. For G-E debris the distribution on the sky is relatively
smooth already after 2 Gyr of integration, implying that the debris is fully
phase-mixed. In this context the asymmetries reported by \citet{Iorio19} if
attributed to the merger of G-E, must be interpreted as being due to
incompleteness in the sky distribution. 
\rev{N-body simulations that model this merger which have too short
integration times  \citep{Donlon20}, or which do not include the growth of
a massive (Galactic) disk that brings axial symmetry to the gravitational
potential after the merger is completed \citep[as in,
e.g.][]{Villalobos2008,Naidu21}, predict unevenness in the spatial
distribution of the distant debris. This unevenness should be apparent in
asymmetries in e.g. the $v_z$ distribution for nearby \mbox{G-E}
stars, which are in fact, not seen in the data \citep[see e.g.][]{Helmer19a}}. On the
other hand, such kinematic asymmetries are seen for the HStr (and manifest
themselves in feebly uneven sky distributions, as shown in the top right panel
of Fig.~\ref{fig:allsky}), and have been used to constrain this accretion event
to have taken place 5-8~Gyr ago \citep{Helmer19b}.  
 
The established associations are also useful for probing different regions
inside the progenitor before this was disrupted \citep{Helmer20, Naidu21}.
Comparisons in terms of stellar populations, their ages and chemistry could for
example constrain the presence of gradients in the parent systems, and in this
way, determine the characteristics of the building blocks at the time of
accretion. 

\begin{acknowledgements}
We thank the anonymous referee for a constructive and useful report. We
acknowledge support from a Vici grant and a Spinoza prize from the Netherlands
Organisation for Scientific Research (NWO). We are also grateful to Thijs
M\"ollenkramer for first explorations that put this work on firmer grounds. 

We have made use of data from the European Space Agency (ESA) mission \gaia
(https://www.cosmos.esa.int/gaia), processed by the \gaia Data Processing and
Analysis Consortium (DPAC,
https://www.cosmos.esa.int/web/gaia/dpac/consortium).  Funding for the DPAC has
been provided by national institutions, in particular the institutions
participating in the \gaia Multilateral Agreement.

The following software packages where used in this publication:
    \package{Astropy} \citep{astropy, astropy:2018},
    \package{healpy} \citep{healpix, healpy},
    \package{IPython} \citep{ipython},
    \package{matplotlib} \citep{mpl},
    \package{numpy} \citep{numpy},
    \package{scipy} \citep{scipy},
    \package{vaex} \citep{vaex}
\end{acknowledgements}


\bibliographystyle{aa} 
\bibliography{refs.bib} 
\end{document}